# Bringing Leaders of Network Sub-Groups Closer Together Does Not Facilitate Consensus

Matthew I. Jones[1,2,3*] and Nicholas A. Christakis[1,2,4]

Affiliations
[1] Yale Institute for Network Science, Yale University; New Haven, CT, USA.
[2] Department of Sociology, Yale University; New Haven, CT, USA.
[3] Sunwater Institute, North Bethesda, MD, USA.
[4] Department of Statistics and Data Science, Yale University; New Haven, CT, USA.
* Corresponding author email: matt.jones@yale.edu

## Abstract

Consensus formation is a complex process, particularly in networked groups. When individuals are incentivized to dig in and refuse to compromise, leaders may be essential to guiding the group to consensus. Specifically, the relative geodesic position of leaders (which we use as a proxy for ease of communication between leaders) could be important for reaching consensus. Additionally, groups searching for consensus can be confounded by noisy signals in which individuals are given false information about the actions of their fellow group members. We tested the effects of the geodesic distance between leaders (geodesic distance ranging from 1-4) and of noise (noise levels at 0%, 5%, and 10%) by recruiting participants (N=3,456) for a set of experiments (n=216 groups). We find that noise makes groups less likely to reach consensus, and the groups that do reach consensus take longer to find it. We find that leadership changes the behavior of both leaders and followers in important ways (for instance, being labeled a leader makes people more likely to 'go with the flow'). However, we find no evidence that the distance between leaders is a significant factor in the probability of reaching consensus. While other network properties of leaders undoubtedly impact consensus formation, the distance between leaders in network sub-groups appears not to matter.

## Introduction

Consensus formation is a critical process in social systems. It requires sharing information about personal preferences with members of the group, processing information received from others, and then perhaps updating preferences or making concessions to align with the rest of the group. Consensus is complex, requiring communication[1,2], trust[3], suppression of bias[4], and strong incentives[5], and it appears in many aspects of society, including domestic politics[6], international treaties[7,8], and jury decisions[9]. Because of this ubiquity[10], consensus has been studied theoretically [11,12] and empirically[5,13]. Diverse conditions[14], individual-level behaviors[15], and group-level policies [16] are important factors that can foster or suppress consensus.

While searching for consensus, the spread of information through a group frequently occurs on a communication network which connects individuals. It has been well established that when group behavior occurs on a network, the structure of that network has an impact on the behavior of the group[17-21]. Certain networks are more conducive to information transfer than others, which can have an impact on the ability of the group to reach a consensus[22]. For example, strong community structure can slow consensus formation in experimental groups[23].

Separately, leadership is also important to group performance. Good leaders can delegate tasks to followers so the group functions as a single, cohesive unit[24]. They can act as information clearinghouses, providing a central hub for information to be collected and distributed in an organized manner[25]. They can wield authority to bring unruly group members' behavior in line with the rest of the group[26,27]. And they can speed up difficult decisions by executive action, reducing the need for lengthy deliberation and discussion with decisive fiat[28].

All these features of leadership suggest that when a group is searching for consensus, the actions of leaders should be critical for success, particularly in a networked setting with locally contiguous information flow. To test this hypothesis, we designed an experiment to study the intersection of consensus, networks, and leadership. We first divide the group into two inter-connected factions with differing preferences but with the common ultimate goal of reaching consensus. By providing each faction with a leader and asking the entire group to reach consensus, we can determine the value of leaders, how they affect individual behavior, and what effect, if any, they have on the success of the sub-group factions or the group as a whole, in a variety of conditions. When considering the leaders of potentially opposing factions on a network, it is natural to consider the geodesic distance between leaders, which can serve as a proxy for ease of communication between the two competing factions[29].

Our experiment models two groups reaching consensus on a contentious issue. We recruited N=3,456 human participants (from the online labor market, Prolific) to take part in live online tasks. In each experiment, 16 participants were placed in a network and asked to reach consensus on one of two colors. However, the group was divided into two teams of eight, each with a preference for one of the colors. Each team was also given a leader (using a minimal leadership paradigm[30]), and the relative positions of the leaders was manipulated to experimentally study the effect of the geodesic separation of the leaders on group consensus.

We also varied the difficulty of the consensus problem by adding noise. In our experiment, with some probability, participants could be given incorrect information about the color choice of one or more of their neighbors. In certain tasks, noise is essentially misinformation, a distraction that makes a task more difficult[31], while in other tasks, it has been shown empirically [32] and theoretically [20] that noise can actually help to break up gridlock so as to reach optimal solutions.

## Results

**The Consensus Game**

We recruited participants online to take part in experiments with the open-source Breadboard software [33](https://breadboard.yale.edu/), which we have used to create a coloring game somewhat similar to previous experiments [5,23,32] that ask participants to solve a global optimization problem using only local information. Each game required 16 participants to take part.

The goal of the game – to unanimously agree on either red or blue – is completed the moment all participants have selected the same color. Participants can freely change their color between red and blue as the game progresses with two buttons on their screen. Their current color choice is shown in real time to each participant's neighbors.

At the center of our experiment is the network we use for the consensus game, shown in Figure 1. Strong community structure divides the network into two factions, indicated by the square and circle shapes. Members of the two factions are incentivized to prefer different colors (see below for a description of participant payoffs). Leaders for both factions are identified by a gold border, and the geodesic distance between leaders can take any value between 1 and 4. We experimentally manipulated the square leader position by moving it to any vertex with an indicated number in Figure 1. Critically, the four potential square leader vertices are isomorphic; they occupy identical positions in the network that are only distinguished by their distance from the blue leader. In fact, all vertices in the grey rectangle are isomorphic (Type A vertices), and all vertices not in the grey rectangle are also isomorphic (Type B vertices). When analyzing individual behavior, we are interested in three group of vertices: the two leaders, the six non-leaders who see their leader, and the four Type A non-leaders who do not see their leader. Therefore, the four Type B non-leaders who do not see their leader do not appear in our individual-level analysis.

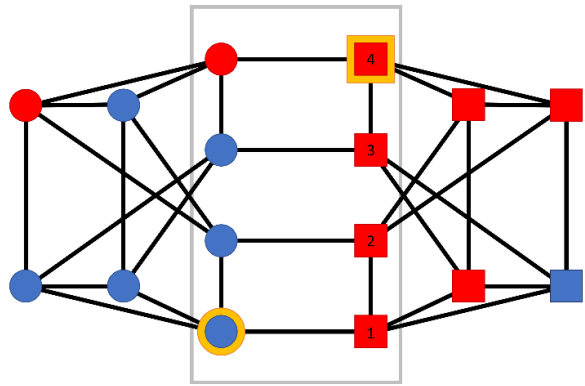

*Figure 1: The network used for our experiment. The two factions with different preferences are indicated by shape and Type A vertices are located inside the grey rectangle (otherwise they are type B). The leaders are indicated by gold borders, and the numbered red vertices are potential leader positions with different leader geodesic distances. In this snapshot of the experiment, where the leaders are 4 edges apart, 9 vertices have chosen red and 7 have chosen blue.*

In Figure 1, participants in the circle group are incentivized by earning a larger bonus payment if the final color is (say) blue, while participants in the square group earn more if the final color is (say) red (the relationship between shape and preferred color is randomized for each task; see Methods). Participants earn $1.25 if the entire group agrees on their preferred color, while the other group earns $0.75.  If the group cannot decide on a color within five minutes, the game ends and the participants do not earn any bonus. An additional bonus of $1 gradually decreases over the course of the game to encourage reaching a solution quickly; for example, if the group finishes with half of the time remaining, each participant earns an additional $0.50.

The leaders are randomly selected from the participant pool. They are leaders in name only; they have no more information than the other participants; they have no additional abilities or responsibilities (except not being directly exposed to noise); and they do not have uniquely central positions in the network. In fact, they have only three in-group neighbors, while the Type B vertices in the group have four. All participants are informed about the existence of leaders even if they cannot see either faction leader. In other words, the treatment here is to simply label one node as the "leader," to inform that person of this label (thereby possibly modifying their own behavior), to show those to whom this leader is connected that one of their alters is a leader (perhaps modifying their behavior), and to inform all participants that leaders exist in the groups. The experimental manipulation here is to vary the geodesic distance between the leaders of the two factions.

We also introduce noise into the system. Noise can make a task easier or harder, depending on the specifics of the task. Every 15 seconds, each directed edge that does not lead to or from a leader is treated as "noisy" with some probability p. If an edge from vertex V to vertex U is noisy, then V will see the incorrect color for U. If U has chosen red, then it will falsely appear to V that U is blue. We use three values for p. When p=0, there is no noise, and all information is conveyed accurately; when p=0.05, we expect 2.4 directed edges to be noisy at any given moment; and when p=0.1, there should be about 5 noisy edges. Participants are not told when an edge is noisy, or even that it is possible for edges to transmit inaccurate information. All edges are updated for noise simultaneously, but because the probability of noise is so low, it is rare that participants will see two or more of their neighbors' colors change at the exact same time.

Our final experiment used a factorial design, with 4 values of leadership distance and 3 values of noise. For each cell, we recruited 18 groups of 16 to attempt to reach consensus. The trajectory for each of these groups as the task takes place are in Figure 2. The y-axis, showing the fraction of participants choosing red, represents the level of consensus. Once a group reaches either 0 or 1, consensus is reached and the task ends. The average trajectory of each cell is shown in red. We also report the number of groups that reached consensus in each cell in Table 1.

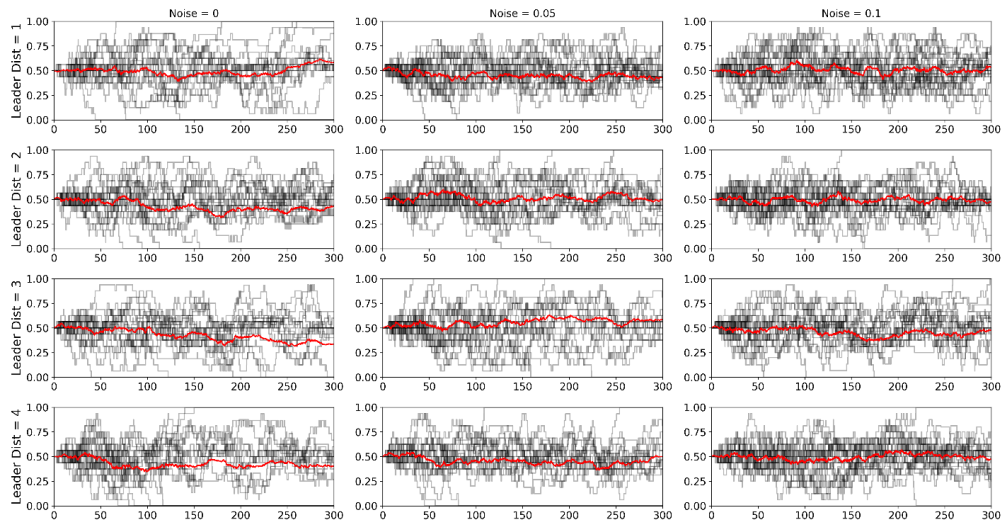

*Figure 2: Trajectory curves for all 4\*3\*18=216 experiments. The columns represent different values of noise and the rows are different distances between leaders. In each plot, 18 curves show the fraction of the population that is choosing red at any time, and the average is shown in red. The x-axis is time in seconds.*

|  | Noise = 0 | Noise = 0.05 | Noise = 0.1 |
|---|---|---|---|
| Distance between leaders = 1 | 7/18 | 2/18 | 0/18 |
| Distance between leaders = 2 | 4/18 | 5/18 | 2/18 |
| Distance between leaders = 3 | 6/18 | 6/18 | 5/18 |
| Distance between leaders = 4 | 7/18 | 5/18 | 2/18 |

*Table 1: The probability that a group successfully reached consensus for each noise and leader distance condition.*

**Minimal Leadership Paradigm Affects Individual Behavior**

When searching for consensus, players would frequently switch between colors, even if nothing changed among their neighbors. We take a probabilistic approach to individual behavior to study how environment affects behavior. For any individual, we can determine the probability that, at a random time during the task, they are playing their non-preferred color. We refine our analysis by conditioning this probability on the number of the ego's neighbors that are playing the ego's non-preferred color and the ego's position in the network. Our dataset looks at 2,592 different individuals (we ignore Type B vertices that do not see their leader) who may face different conditions (e.g. a leader with two neighbors playing their non-preferred color or a non-leader with three neighbors playing their non-preferred color, one of whom is the individual's leader – see Figure 3 for the 18 such conditions), for a total of $n = 12,707$ observations. To account for the fact that the same individual can appear in more than one condition, we use individual-level random effects, and we also account for session-level random effects (see Methods). As shown in Figure 3, regardless of an individual's situation, they are more likely to have their non-preferred color selected when more of their neighbors are already playing the non-preferred color.

Next, to see how leadership affects behavior, we consider four different types of players. First, to see how leaders affect those around them, we look at vertices who can see their leader and compare their behavior when the leader is playing their preferred color to their behavior when the leader is playing their non-preferred color. These probabilities are represented by the orange and green bars in Figure 3. Of course, there is no green bar in the leftmost column, because the leader would simultaneously be playing the preferred and non-preferred color, and therefore there is no data. For the same reason, there is no orange bar in the rightmost column. In Figure 3, we see consistent evidence that a leader playing a non-preferred color makes it more likely that an ego plays their non-preferred color than if a non-leader plays that color. When the number of neighbors playing the non-preferred color is 1, 2, or 3, we can use a Wald test to test the hypothesis that the difference of the "Leader Pref" (the leader is playing the preferred color of their group) and "Leader Non-Pref" (the leader is playing the non-preferred color) coefficients is zero. When the number of neighbors is 1, for example, the coefficients for Leader Pref and Non-Pref are 0.133 and 0.214, respectively. The test has a $\chi^2$ value of 61.173 and a $p$ value of $5.2e-15$, so we conclude that the coefficients are significantly different. Similarly, when the number of neighbors is 2, coefficients are 0.302 and 0.390 ($\chi^2 = 77.37$, $p < 2.2e-16$), and when number of neighbors is 3, coefficients are 0.449 and 0.574 ($\chi^2 = 93.824$, $p < 2.2e-16$).

We are also interested in how the mantle of leadership (as minimal as it is in this experiment) affects behavior. In Figure 3, we can compare leaders (cyan bars) to non-leaders in Type A vertices who do not see their leader (purple bars). These two types of players occupy identical positions in the network and have identical types of neighbors, but we see that they can behave differently. When only a few neighbors are playing the nonpreferred color, leaders and non-leaders are indistinguishable. However, when three or four neighbors are playing the non-preferred color, the leaders are more likely than non-leaders to play their non-preferred color. When three neighbors are playing the non-preferred color, the coefficients are 0.567 for leaders and 0.474 for non-leaders; a Wald test confirms that the difference of these coefficients is not zero ($\chi^2 = 26.084, p = 3.3e-07$). Likewise, the coefficients for four neighbors is significantly larger for leaders (0.772) than for non-leaders (0.614) ($\chi^2 = 46.954, p = 7.3e-12$). In short, being a leader makes one more likely to "go with the flow" and follow the will of the majority. The leaders in this task seem to provide stability, reinforcing to their subordinates that the non-preferred color is the better option, rather than trying to drive change and coerce their followers into compromising on color choice.

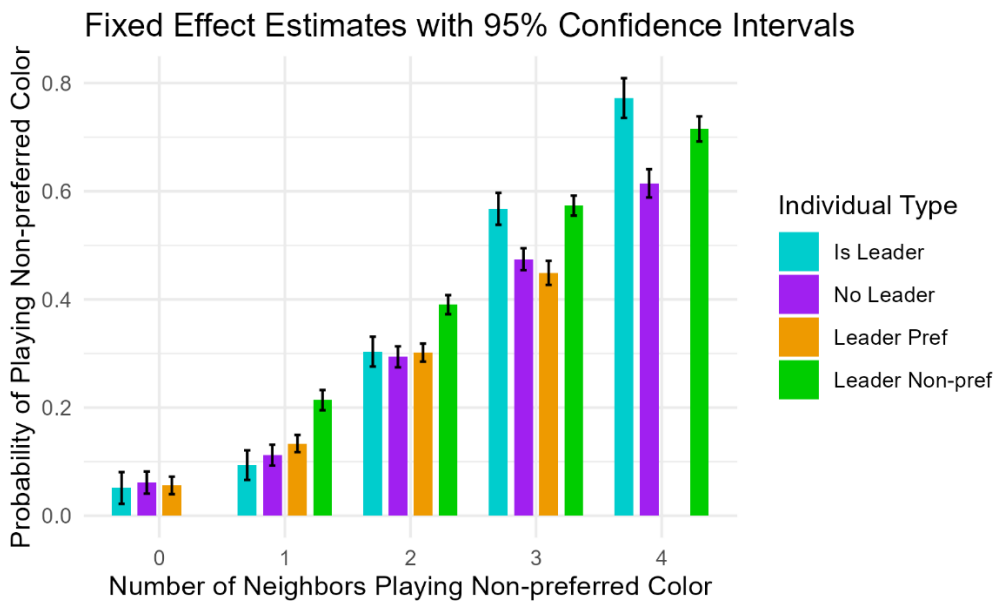

*Figure 3: The probability that individuals in different conditions have selected their non-preferred color at any given time. The "Is Leader" and "No leader" categories contain leaders or non-leaders in Type A vertices that do not see their leader. By comparing these, we see significant differences when 3 or 4 neighbors are playing the ego's non-preferred color. The "Leader Pref" and "Leader Non-Pref" categories contain non-leaders who see their leader, and compare the probability of selecting the non-preferred color when their leader has selected the preferred or non-preferred color. Here, we see that the leader playing the non-preferred color has a larger effect than a non-leader on the ego's color choice.*

**Adding Noise Makes Consensus Difficult**

The introduction of noise can counterintuitively help solve optimization problems by encouraging non-optimal moves that escape local optima and ultimately lead to the global optimal solution[32]. At first glance, the present task seems like it could also benefit from noise. Since all individuals begin with their preferred color and the biased payoffs reward hardliners who refuse to play their non-preferred color, it is plausible that noise could trick participants into compromising and ultimately reaching consensus. However, we see from Figure 3 that many participants already understand the necessity of compromise, playing their non-preferred color over 10% of the time when only one neighbor is playing the non-preferred color.

When it is indeed not needed to break up gridlock, noise can function more to distract and mislead, which is what we see here. The noisy edges that spread false information about neighbors' color selections prevent participants from choosing the color that would ultimately lead to consensus. When we estimate a regression model (see Methods) regarding the frequency of consensus (the values in Table 1), we find a statistically significant negative effect of noise on consensus formation ($\beta_2 = -2.083, p = 0.003$).

In fact, not only does noise reduce the chances of successfully reaching consensus, it also makes the groups that do reach consensus take longer to do so. Using Cox proportional hazard models[34], we analyzed the effect of noise (and leadership position) on the time it takes to reach consensus, and find a significant negative effect (hazard ratio $= 2.670e - 05, p = 0.0034$). In the presence of noise, consensus is less likely and it also takes longer to reach.

**Leader Position Does Not Affect Group-Level Consensus**

Despite the established effects of leadership on the behavior of leaders and followers, we find no evidence in the regression model that the relative position of leaders has an effect on the probability of a group as a whole reaching consensus ($\beta_1 = 0.03889, p = 0.13$). We also find no effect on time to consensus in the proportional hazard model (hazard ratio $= 1.212, p = 0.13$). And including an interaction term also reveals no significant interaction between noise and distance (see Methods). However, perhaps surprisingly, although *not* statistically significant, the coefficients for distance on both consensus frequency and time to consensus are positive. This indicates that groups are more likely to reach consensus and reach consensus faster if leaders are farther apart, each separately affecting their own fiefdoms rather than being close together in constant communication.

## Discussion

The geodesic distance between two vertices in a network is a good proxy for the ability of these vertices to communicate and coordinate. Leaders have disproportionate influence in the coordination game, but their ability to reinforce coordination seems to be irrelevant to the success of the group in these experiments. Rather, they appear to provide a stabilizing effect on their neighbors, keeping them from randomly changing colors and hence preventing consensus. This hints at why leaders being farther apart might actually improve group performance. When the leaders are adjacent, they have no influence over the other side of the whole network, which prevents the entire group from reaching consensus.

However, these results should not suggest that other aspects of leadership are irrelevant to group consensus. Network properties of leaders other than their geodesic separation may still play a large role in the success of the group. Recall that leader degree and centrality was fixed even as leader distance changed in these experiments; increasing these properties for leaders will certainly expand their influence in the group in ways that our experiment did not and may lead to other consensus results that depend on social contagion[35]. Further investigation into leaders' network properties during consensus experiments may shed light on the specific network statistics that are important for leadership to influence consensus formation.

It is also important to recognize that these results are limited by the design of the experiment. Several key factors prevent us from drawing stronger conclusions about the irrelevance of leader distance in consensus. This experiment is highly stylized. Participants are not able to communicate in any way other than selecting a color which is shown to their neighbors. They are not able to persuade or bargain with each other. Similarly, real leaders may use charisma and likeability to convince others to follow them, but our leaders had only the minimal label of leader, represented by a gold border, and no other credentials to persuade others. They were also chosen randomly from the group, instead of selecting those with personalities that lend themselves to leadership. Despite these drawbacks, we still see strong evidence of the effect of leadership on individual behavior. Critically, we confirm previous research[36] claiming that good leaders follow.

As one illustration of the potential relevance of our findings, consider the example of how the U.S. Congress is broken up into domain-specific committees such as Agriculture, Budget, and Homeland Security. Folk wisdom is that one of the key factors in the health of a committee (measured by the amount of legislation passed, the number of hearings held, and so on) is the relationship between the Committee Chair and the Ranking Minority Member[37-39]. These two individuals function in similar ways to the leaders in our experiment, each heading up a team with opposing preferences but ostensibly trying to find compromise and common ground. Our findings indicate that the relationship between these two leaders may not be as important as the anecdotes suggest. Instead, we suspect that other network properties or behaviors of the leaders (which could of course be correlated with leader geodesic distance) may be more important for facilitating consensus.

Unlike in anti-coordination activities like graph coloring, gridlock is less of a problem in consensus coordination problems and noise is detrimental to the success of the group. Perhaps humans have evolved or learned to be "naturally noisy" in coordination problems but not in anti-coordination problems, or perhaps anti-coordination problems are naturally more difficult due to the network structures on which they are solved. Future work could examine this relationship in more detail.

Leaders behave differently and have special influence in their own sub-groups, but, despite this, the relationship among leaders of competing groups – in the sense of their geodesic separation at least – may not be a factor in reaching overall agreement. Reaching a broad consensus in groups is not easy.

## Methods

This study was approved by the Yale University Committee on the Use of Human Subjects. All subjects gave their informed consent in accordance with the Yale University IRB before participating in the study.

This study was preregistered at https://osf.io/n3qa2/. The main effects of noise and the distance between leaders on consensus frequency and speed were preregistered. All individual-level analyses were exploratory. The order of data collection was randomized to avoid order effects.

Participants were recruited using Prolific (https://www.prolific.com/), an online recruitment service. Upon joining the study, participants are redirected to our Breadboard server, where they must first consent to participating in our study and then read through a short tutorial and pass a comprehension quiz. Per Prolific policy, participants are given two attempts to pass the 2-question, multiple choice quiz. Also, immediately before the experiment begins, participants must pass an additional attention check within 30 seconds to ensure that they are prepared to participate in the live task. Participants that complete the attention check earn a base pay of $4.50 and are eligible to participate in the rest of the experiment. Participants are given 5-10 minutes to read through the tutorial and pass the quiz before the experiment begins.

To ensure that we recruit enough participants to run the experiment, we typically recruited more than the necessary 16 participants. Excess participants were randomly chosen and dismissed from the experiment with the base pay, and the remaining 16 participants began playing the consensus game.

During the consensus game, participants must select a color within every minute to avoid being dropped for inactivity. At 30 seconds after their last color choice, a 30 second timer appears warning them that they are about to be dropped from the experiment. Players can select the same color to avoid being dropped without having to change their color. This inactivity timer, a recaptcha verification step (https://www.google.com/recaptcha/about/), the tutorial comprehension quiz, and the pre-experiment attention check serve to filter out bots and other inattentive participants.

During the game, participants were reminded regarding: the remaining time left in the game, their team and preferred color, their payoff incentives, the identification of leaders, and how to change their color. An example of their screen during the game is shown below.

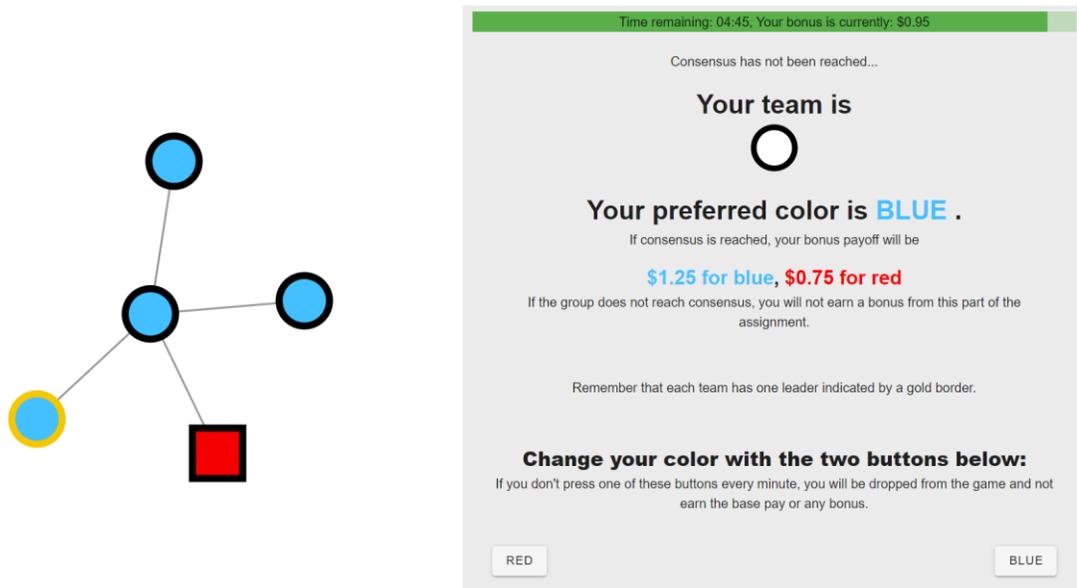

*Figure 4: Players' screens during the game. The colors of their neighbors were updated in real time as color changes were made.*

Participants that are dropped from the coloring game for inactivity could disrupt the behavior of the other participants. Therefore, if a replicate had two non-leaders or one leader drop out before the game ends, that experiment was discarded and the replicate repeated with new participants; in total, 17 replicates needed to be repeated because of dropped participants.

Edges in the network are undirected, but the presence of noise was directed. That is, player A may be getting incorrect information about player B's color, but player B may still get correct information about player A's color. Every 15 seconds, each directed edge that is not connected to a leader is treated as "noisy" with the desired probability. Participants are not told when an edge is noisy, or even that it is possible for edges to transmit inaccurate information.

**Group-level Statistical Analysis**

We examined the effect of leader distance and noise on the probability of the group reaching consensus. We estimated the following regression model

$$p_i = \beta_0 + \beta_1 * D_i + \beta_2 * N_i + \epsilon$$

where $p_i$ is the probability of group $i$ reaching consensus, $D_i$ is the distance between leaders in group $i$, $N_i$ is the level of noise in group $i$, and $\epsilon$ is the error to be minimized.

Specifying the following model which includes an interaction term between leader distance and noise did not reveal any additional results.

$$p_i = \beta_0 + \beta_1 * D_i + \beta_2 * N_i + \beta_3 * D_i * N_i + \epsilon$$

We also examined the effect of distance and noise on the time it takes to reach consensus. A hazard model approximates the risk of an event (in this case, reaching consensus) occurring in a short span of time, denoted by $\lambda$. In our model, a group's hazard is dependent on both leader distance and noise according to

$$\lambda(t|X_i) = \lambda_0(t) \exp(\beta_4 * D_i + \beta_5 * N_i)$$

The hazard ratio of a factor is then $\exp(\beta_j)$.

**Individual-level Statistical Analysis**

For our analysis of individual behavior, we use a mixed-effects model with random effects accounting for both individual id and session id and fixed effects for the individual's condition. There are four types of individuals we look at: is a leader, cannot see leader (but is Type A vertex), sees their leader when leader is playing the preferred color, and sees their leader when leader is playing the non-preferred color. Specifically, for an individual $i$ in session $j$ of type $T$, the probability of playing their non-preferred color is

$$p_{i,j,T} = \gamma_i + \gamma_j + \beta_T$$

This model allows us to estimate the effect of neighbors' behavior on individual behavior (the $\beta$ coefficients which are plotted in Figure 3) while also accounting for individual variation.

Statistical analyses were done in R v4.4.1[40].

**Acknowledgments**

Wyatt Israel provided expert programming assistance. This work was supported by the Sunwater Institute and the Paul Graham Foundation (NAC).

**Author Contributions Statement**

Conceptualization, Methodology and Writing: MJ and NAC; Software, Data Collection, and Statistical Analysis: MJ; Funding Acquisition and Supervision: NAC.

**Competing Interests Statement**

The authors declare that they have no competing interests.

**Data Availability Statement**

The data used in this study have been made publicly available at [UPLOADED PRIOR TO PUBLICATION]. During the review process, data will be made available upon request.

# References

1    King, A. J. *et al.* Performance of human groups in social foraging: the role of communication in consensus decision making. *Biol. Lett.* **7**, 237-240 (2010).
2    Garrod, S. & Doherty, G. Conversation, co-ordination and convention: an empirical investigation of how groups establish linguistic conventions. *Cogn.* **53**, 181-215 (1994).
3    Lewicki, R. J. & Stevenson, M. A. Trust Development in Negotiation: Proposed Actions and a Research Agenda. *Bus. Prof. Ethics J.* **16**, 99-132 (1997).
4    Lord, C. G., Ross, L. & Lepper, M. R. Biased assimilation and attitude polarization: The effects of prior theories on subsequently considered evidence. *J. Pers. Soc. Psychol.* **37**, 2098-2109 (1979).
5    Kearns, M., Judd, S., Tan, J. & Wortman, J. Behavioral experiments on biased voting in networks. *Proc. Natl. Adad. Sci. U.S.A.* **106**, 1347-1352 (2009).
6    Krehbiel, K. Unanimous Consent Agreements: Going Along in the Senate. *J. Politics.* **48**, 541-564 (1986).
7    Gifkins, J. Beyond the Veto: Roles in UN Security Council Decision-Making. *Glob. Gov. Rev. Multilater. Int. Organ.* **27**, 1-24 (2021).
8    Shifrinson, J. Time to Consolidate NATO? *Wash. Q.* **40**, 109-123 (2017).
9    Feddersen, T. & Pesendorfer, W. Convicting the Innocent: The Inferiority of Unanimous Jury Verdicts under Strategic Voting. *Am. Political Sci. Rev.* **92**, 23-35 (1998).
10   Baronchelli, A. The emergence of consensus: a primer. *R. Soc. Open Sci.* **5**, 172189 (2018).
11   Degroot, M. H. Reaching a Consensus. *J. Am. Stat. Assoc.* **69**, 118-121 (1974).
12   Jones, M. I., Pauls, S. D. & Fu, F. The dual problems of coordination and anti-coordination on random bipartite graphs. *New J. Phys.* **23**, 113018 (2021).

13  Chu, O. J., Donges, J. F., Robertson, G. B. & Pop-Eleches, G. The microdynamics of spatial polarization: A model and an application to survey data from Ukraine. *Proc. Natl. Acad. Sci. U.S.A.* **118**, e2104194118 (2021).
14  Lu, Q., Korniss, G. & Szymanski, B. K. The Naming Game in social networks: community formation and consensus engineering. *J. Econ. Interact. Coord.* **4**, 221-235 (2009).
15  Boyd, R. & Richerson, P. J. Punishment allows the evolution of cooperation (or anything else) in sizable groups. *Ethol. Sociobiol.* **13**, 171-195 (1992).
16  González-Avella, J. C., Cosenza, M. G. & Tucci, K. Nonequilibrium transition induced by mass media in a model for social influence. *Phys. Rev. E.* **72**, 065102 (2005).
17  Jones, M. I., Pauls, S. D. & Fu, F. Containing misinformation: Modeling spatial games of fake news. *PNAS Nexus* **3**, pgae090 (2024).
18  Lieberman, E., Hauert, C. & Nowak, M. A. Evolutionary dynamics on graphs. *Nature* **433**, 312-316 (2005).
19  Komarova, N. L., Schang, L. M. & Wodarz, D. Patterns of the COVID-19 pandemic spread around the world: exponential versus power laws. *J. R. Soc. Interface.* **17**, 20200518 (2020).
20  Jones, M. I., Pauls, S. D. & Fu, F. Random choices facilitate solutions to collective network coloring problems by artificial agents. *iScience* **24**, 102340 (2021).
21  Shirado, H., Crawford, F. W. & Christakis, N. A. Collective communication and behaviour in response to uncertain 'Danger' in network experiments. *Proc. R. Soc. A.* **476**, 20190685 (2020).
22  Dall'Asta, L., Baronchelli, A., Barrat, A. & Loreto, V. Agreement dynamics on small-world networks. *Europhys. Lett.* **73**, 969-975 (2006).
23  Judd, S., Kearns, M. & Vorobeychik, Y. Behavioral dynamics and influence in networked coloring and consensus. *Proc. Natl. Acad. Sci. U.S.A.* **107**, 14978-14982 (2010).
24  Yukl, G. & Mahsud, R. Why flexible and adaptive leadership is essential. *Consult. Psychol. J.: Pract. Res.* **62**, 81-93 (2010).
25  Komai, M., Stegeman, M. & Hermalin, B. E. Leadership and Information. *Am. Econ. Rev.* **87**, 944-947 (2007).
26  Kosfeld, M. & Rustagi, D. Leader Punishment and Cooperation in Groups: Experimental Field Evidence from Commons Management in Ethiopia. *Am. Econ. Rev.* **105**, 747-783 (2015).
27  Flack, J. C., Girvan, M., de Waal, F. B. M. & Krakauer, D. C. Policing stabilizes construction of social niches in primates. *Nature* **439**, 426-429 (2006).
28  Klieman, A. S. Preparing for the Hour of Need: The National Emergencies Act. *Pres. Stud. Q.* **9**, 47-65 (1979).
29  Arbesman, S. & Christakis, N. Leadership Insularity: A New Measure of Connectivity Between Central Nodes in Networks. *Connect.* **30**, 4-10 (2010).
30  Otten, S. The Minimal Group Paradigm and its maximal impact in research on social categorization. *Curr. Opin. Psychol.* **11**, 85-89 (2016).
31  Arechar, A. A., Dreber, A., Fudenberg, D. & Rand, D. G. "I'm just a soul whose intentions are good": The role of communication in noisy repeated games. *Games Econ. Behav.* **104**, 726-743 (2017).
32  Shirado, H. & Christakis, N. A. Locally noisy autonomous agents improve global human coordination in network experiments. *Nature* **545**, 370-374 (2017).
33  McKnight, M. E. & Christakis, N. A. in *Breadboard: Software for Online Social Experiments* (Yale University, 2016).
34  Cox, D. R. *Analysis of Survival Data*. (Chapman and Hall/CRC, 1984).

35  Arioldi, E. M. & Christakis, N. A. Induction of social contagion for diverse outcomes in structured experiments in isolated villages. *Science* **384**, eadi5147 (2024).
36  Peters, K. & Haslam, A. A. I follow, therefore I lead: A longitudinal study of leader and follower identity and leadership in the marines. *B J. Psychol.* **109**, 708-723 (2018).
37  Kornberg, M. L. *Inside Congressional Committees: Function and Dysfunction in the Legislative Process*. (Columbia University Press, 2023).
38  Alexander, B. *A Social Theory of Congress: Legislative Norms in the Twenty-First Century*. (Lexington Books, 2021).
39  U.S. House Select Committee on the Modernization of Congress. Enhancing Committee Productivity Through Consensus Building. *https://www.govinfo.gov/app/details/CHRG-117hhrg48599/CHRG-117hhrg48599* (2021).
40  R Core Team. *R: A Language and Environment for Statistical Computing*. (R Foundation for Statistical Computing, 2021).